\documentclass[aps,prb,twocolumn,10pt,superscriptaddress]{revtex4-2}
\usepackage{amssymb,bm,graphicx,ulem,hyperref,amsmath}
\usepackage[dvipsnames]{xcolor}

\begin{document}

\author{Igor Y. Chestnov}
\affiliation{School of Science, Westlake University, 18 Shilongshan Road, Hangzhou 310024, Zhejiang Province, China}
\affiliation{Institute of Natural Sciences, Westlake Institute for Advanced Study, 18 Shilongshan Road, Hangzhou 310024, Zhejiang Province, China}
\affiliation{Vladimir State University, Gorkii St. 87, 600000, Vladimir, Russia}

\author{Sergey M. Arakelian}
\affiliation{Vladimir State University, Gorkii St. 87, 600000, Vladimir, Russia}

\author{Alexey V. Kavokin}
\affiliation{School of Science, Westlake University, 18 Shilongshan Road, Hangzhou 310024, Zhejiang Province, China}
\affiliation{Institute of Natural Sciences, Westlake Institute for Advanced Study, 18 Shilongshan Road, Hangzhou 310024, Zhejiang Province, China}
\affiliation{Russian Quantum Center, Skolkovo IC, Bolshoy Bulvar 30, bld. 1, Moscow 121205, Russia}
\affiliation{NTI Center for Quantum Communications, National University of Science and Technology MISiS, Moscow 119049, Russia}

\email{igor\_chestnov@westlake.edu.cn}

\title{Giant synthetic gauge field for spinless microcavity polaritons \\
	in crossed electric and magnetic fields}

\begin{abstract}
The artificial gauge field for electrically neutral exciton polaritons devoid from the polarization degree of freedom can be synthesized by means of applying crossed electric and magnetic fields. The appearance of the gauge potential can be ascribed to the motional (magneto-electric) Stark effect which is responsible for the presence of a linear-in-momentum contribution to the exciton kinetic energy. We study the interplay of this phenomenon with the competing effect which arises from the  Rabi-splitting renormalization due the reduction of the electron-hole overlap for a moving exciton. Accounting for this mechanism is crucial in the structures with the high ratio of Rabi splitting and the exciton binding energy. Besides, we propose an approach which boosts the gauge field in the considered system. It takes advantage of the crossover from the hydrogen-like exciton to the strongly dipole-polarized exciton state at a specific choice of electric and magnetic fields. The strong sensitivity of the exciton energy to the momentum in this regime leads to the large values of the gauge field. We consider the specific example of a GaAs ring-shape polariton Berry phase interferometer and show that the flux of the effective magnetic field may approach the flux quantum value in the considered crossover regime.
\end{abstract}

%\noindent{\it Keywords\/}:{synthetic gauge field, exciton polaritons, Berry phase, quantum wells}

%\submitto{\NJP}
\maketitle

\section{Introduction}

Artificial gauge fields allow  for revealing of many peculiar phenomena such as fractional quantum Hall and Aharonov-Bohm effects, various topological phases of matter  etc{.} in the systems  decoupled from the real gauge fields. The paradigmatic examples here are photonic systems \cite{Ozawa2019,Lu2014}, ultra-cold atomic gases \cite{Galitski2019,Goldman2014,Goldman2016} and solid state media \cite{Si2016}. The synthesis of the artificial gauge fields for electrically neutral particles is aimed at making them to behave as if they are charged and affected by the real electromagnetic field. Synthetic gauge fields can be described by an effective vector potential $\mathbf{A}_{\rm eff}$  that results is the appearance of an effective magnetic field $\mathbf{B_{\rm eff}}=\bm{\nabla}\times\mathbf{A_{\rm eff}}$. The Schr\"oedinger equation for a moving particle of a charge $q$ is modified in this case by a substitution $\hat{\mathbf{p}} \rightarrow \hat{\mathbf{p}} - q\mathbf{A}_{\rm eff}$. 
Therefore, in the case of a parabolic dispersion $\hat{H} \propto \hat{\mathbf{p}}^2$, the gauge field is manifested by  the presence of the term linear in $\mathbf{p} \mathbf{A}_{\rm eff}$ in the single-particle dispersion relation.

\begin{figure}
	\includegraphics[width=\linewidth]{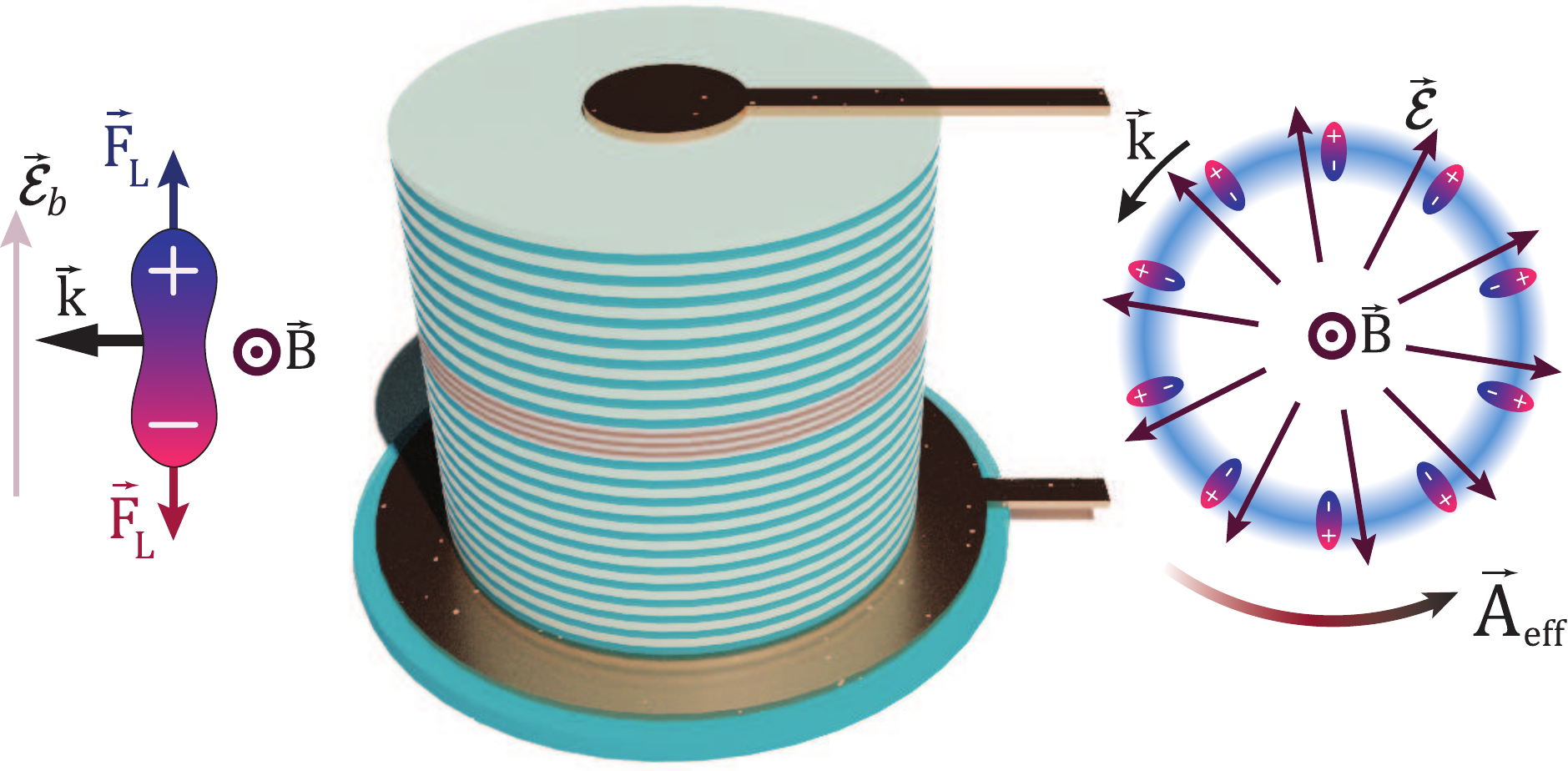}
	\caption{The sketch of a micropillar cavity with a ring-shaped polariton condensate excited by a nonresonant optical pump. The electrodes generate an electric field  with the divergent in-plane component. In combination with the magnetic field oriented along the pillar growth axis, it generates the azimuthal effective gauge potential for polaritons \cite{Imamoglu1996,Xue2019}. The coupling between the collective and the internal motions of the exciton which is responsible for the effective electric field appearance is illustrated on the left. 
}\label{fig1}
\end{figure}

Recently, the synthesis of the tunable artificial gauge field for electrically neutral superfluid of exciton polaritons was demonstrated \cite{Lim2017}. Microcavity polaritons are hybrid quasiparticles which arise from the coherent coupling of photons with the quantum well excitons. The  approach employed in \cite{Lim2017} takes advantage of the excitonic magneto-electric Stark effect  \cite{Thomas1961} which occurs in crossed electric and magnetic fields. The electric field endows exciton with the static dipole moment $\mathbf{d}$. If the exciton moves in perpendicular magnetic field, it feels the effective electric field $\bm{\mathcal{E}}_b$ which originates from the oppositely directed Lorentz forces acting on the electron and the hole, see Fig.~\ref{fig1}. The strength of this field is governed by the exciton center-of-mass momentum, $\bm{\mathcal{E}}_b = \mathbf{p} \times \mathbf{B}/M$, where $M$ is the exciton mass. Therefore the moving dipole-polarized exciton interacts with the effective electric field with the energy $-e  \mathbf{d}  \cdot \bm{\mathcal{E}}_b = \mathbf{d}\times \mathbf{B}\cdot \mathbf{p}/M$ which is linear in $\mathbf{p}$. It leads to the appearance of the effective gauge potential which affects polariton dynamics via their excitonic component.

In the experiment  \cite{Lim2017}, the presence of the gauge field was evidenced by a non-reciprocal polariton transport  which can be connected with different geometrical phases accumulated along and opposite to the gauge field directions. The origin of these phases can be attributed with He-McKellar-Wilkens (HMW) effect \cite{He1993,Wilkens1994,Wei1995}. It predicts the acquisition of the Berry phase by the electric dipole moving in the magnetic field. This effect possesses a fundamental similarity  with the Aharonov-Bohm and the Aharonov-Casher effects which are responsible for the relevant phenomena for a charge moving in  a magnetic field and for a magnetic dipole (spin) moving in an electric field, respectively \cite{Chen2013}.

The nontrivial manifestations of the HMW effect  were initially foreseen in the closed geometry where the Berry phases acquired by the dipoles travelling adiabatically along different trajectories lead to a quantum interference. Such phenomena were predicted in a superfluid helium and dipolar condensate localized on a torus. Generation of the HMW phase in this case implies the use of either radial electric and axial magnetic fields \cite{Sato2009,Wei1995,Wood2016} or co-axial electric and radial magnetic fields \cite{He1993}. The latter configuration can also be  realized with no need of electric field once the excitonic superfluid formed in the bilayer systems is considered \cite{Balatsky2004,Sonin2009,Sonin2010}. {The typical manifestation of the quantum interference is an appearance of a circular persistent current of the dipolar superfluid as the Berry phase accumulated along the circumference exceeds $\pi$ \cite{Wood2016}.}

The use of microcavity polaritons instead of the helium or excitonic systems as proposed in \cite{Lim2017} is advantageous because of several aspects. Polariton condensates can be generated at high temperatures and are convenient for the optical detection and control. Besides, as polaritons are partially photons, the presence of the HMW phase paves the way for the realization of the fast tunable gauge field for photons. 

{Moreover, polaritons are subject to strong spin–orbit coupling effects which are a source of the non-Abelian gauge fields \cite{Tercas2014}. The theoretical proposal of a polariton Berry phase interferometer by Shelykh et al. \cite{Shelykh2009} has induced a significant experimental effort towards observation of topological effects in polaritonics. Some of these works have been crowned by a stunning success, the polariton topological insulator \cite{Klembt2018} and the quantum geometric tensor \cite{Gianfrate2020} have been demonstrated. However, the realisation of Shelykh proposal for a ring-shape Berry phase interferometer is still pending. One of the main reason is the pinning of polarization of polariton condensates in laterally confined structures that suppresses the Berry phase originated from the TE-TM splitting of exciton-polariton modes. The recent works on superfluid circular currents of exciton polaritons in micropillars did not reveal any significant polarizations effects \cite{Sedov2020}. In this context, it is important to study the topological effects in a scalar (in contrast to spinor) polariton liquid that is devoid of the polarization degree of freedom. In this work, we discuss the use of HMW effect for the synthesis of the $U(1)$ gauge field for polaritons.}

The generation of the HMW phase with exciton polaritons is sophisticated due to their hybrid nature. The energy of microcavity polations is strongly affected by the strength of the coupling between the excitons and photons known as a Rabi splitting. The effective electric field $\bm{\mathcal{E}}_b$ felt by the moving exciton modifies its internal structure and thus alters the optical oscillator strength and the Rabi splitting as a consequence. In this paper we demonstrate that this effect can be also  used for generation of the artificial gauge field for polaritons. However it turns out that this field is oriented oppositely to the one arising from the HMW effect. Therefore, accounting for both mechanisms is crucial. We study the interplay of theses effects and discuss the ways to maximize the effective gauge field.

The strength of the effective gauge field is the key characteristic of any method of synthesis. The quantum interference effects in the circular geometry \ref{fig1} are expected to occur once the Berry phase acquired along the ring circumference  $\phi=\hbar^{-1}\oint \mathbf{A_{\rm eff}}\cdot d\bm{l}$ is about $\pi$. However  due to the small effective mass of exciton polaritons the gauge field arising due to the motion Stark effect is rather weak. In this paper we propose the approach capable of increasing the gauge field strength by at least an order of magnitude with no need of extremely strong electric or magnetic fields. It implies an operation in the regime of a crossover from the hydrogen-like exciton to the strongly decentred exciton state.

\begin{figure}
	\includegraphics[width=1\linewidth]{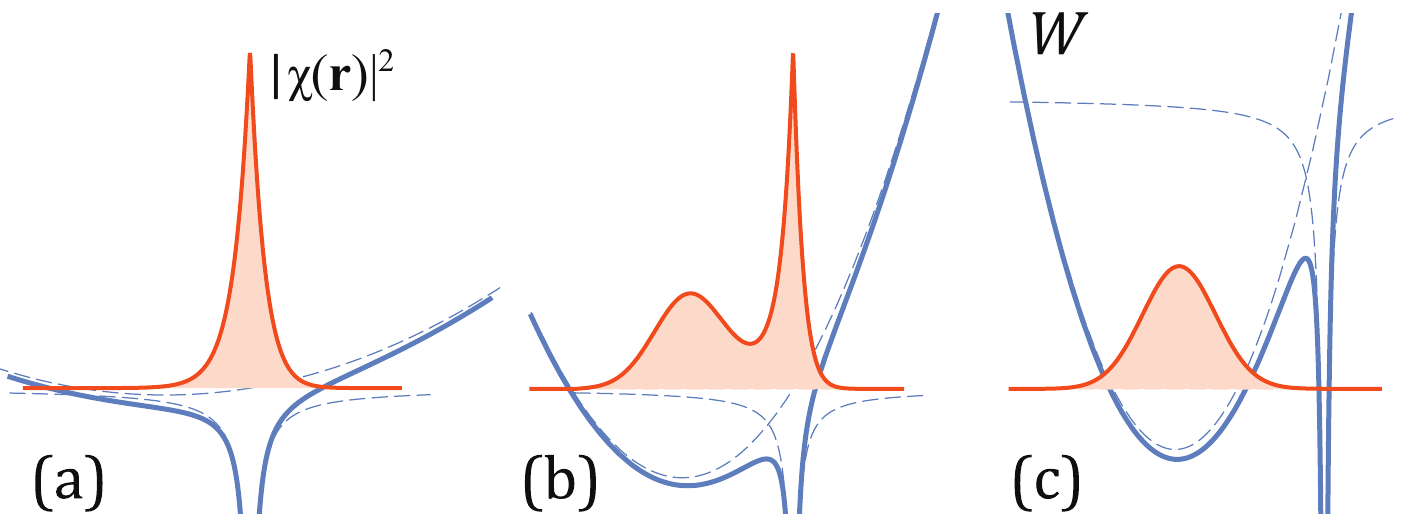}
	\caption{The schematic of a double-well potential $W(\mathbf{r})$ for the 2D electron-hole pair in crossed magnetic and electric fields. (a) The hydrogen-like exciton at weak in-plane electric and normal magnetic field. (b) The case of the moderate electric and magnetic fields illustrating a crossover manifested in the overlap of the wave functions localized in the parabolic diamagnetic potential and in the Coulomb well (dashed curves). The diamagnetic potential is shifted with respect to the Coulomb well due to the presence of electric field. 
	(c) The polarized state possessing large electric dipole at strong magnetic and electric fields.}\label{Fig.States}
\end{figure}

The existence of this crossover is relevant to the problem of two oppositely charged particles placed in crossed electric and magnetic fields which was addressed in \cite{Burkova1976,Escobar2014,Monozon1999,Vincke1992}. 
These studies predict the existence of two types of bound states of the interacting electron-hole pair. The states of the first type correspond to the hydrogen-like exciton (HLE) whose electron average position remains close to the hole so the relative motion wave function $\chi(\mathbf{r})$ is essentially centred in the Coulomb potential well, see Fig.~\ref{Fig.States}a. This is the regime employed for the gauge field synthesis in Ref. \cite{Lim2017}.

For the states of the second kind, the wave function $\chi(\mathbf{r})$  is mostly concentrated outside the Coulomb well. The electron motion is strongly decentred from the hole which is why these states are referred to as decentred.
This is the case of strong electric and magnetic fields when the exciton structure is dominated by the diamagnetic potential whose position is shifted apart from the Coulomb well due to the presence of electric field, see Fig.~\ref{Fig.States}c. These states posses large electric dipole and thus potentially can feel a strong effective gauge field. However due to the vanishing electron-hole overlap they are decoupled from the microcavity photons and can not form polaritons.

Here we focus on the intermediate regime illustrated in Fig.~\ref{Fig.States}b, where a crossover between the states with the weak and strong decentring  occurs \cite{Vincke1992,Lozovik2002}. In the crossover regime, the exciton combines a large electric dipole with the non-vanishing coupling to a photon field. These intermediate states appear to be strongly sensitive to the variation of the electric field strength. Therefore, even weak apparent electric field $\bm{\mathcal{E}}_b$ felt by the moving exciton leads to the significant  modifications of the exciton energy and structure. It causes a significant enhancement of the  effective gauge field strength as compared to the case of the HLE \cite{Lim2017}. 

The paper is organized as follows. In Sec.~\ref{Sec.HLEGaugeField} we evaluate the effective gauge field  for polaritons subjected the crossed electric and magnetic fields. Using a hydrogen-like variational wave function for the  ground state exciton we study the interplay between the HMW effect and the Rabi-splitting renormalization for a moving polariton. This section is summarized by the discussion of the strength of the gauge field  attainable in this regime. Section \ref{Sec.III} presents the study of the artificial gauge field for polaritons formed by non-hydrogenic excitons.  Solving the electron-hole bound state problem numerically, we investigate the properties of the mixed exciton states arising in the crossover regime. The synthesis of the artificial gauge field for polaritons in this regime is discussed in Sec.~\ref{Sec.IV}. The results are summarized in the Conclusions.

\section{Hydrogen-like exciton in the crossed electric and magnetic fields}\label{Sec.HLEGaugeField}
 
The quantum interference effects in the polariton superfluid are expected to occur in the ring-shaped polariton condensates affected by the perpendicular magnetic and radial electric fields \cite{Imamoglu1996,Xue2019}. Such a configuration can be realized in the annual optical traps \cite{Askitopoulos2018} or in a pillar microcavity \cite{Sedov2020}. Alternatively one can think about constructing of the Berry phase interferometer in the design proposed in \cite{Shelykh2009}.	
	
To be specific, we consider a ring-shaped condensate of exciton polaritons excited in the configuration shown in Fig.~\ref{fig1}. A combination of a spot-like inner and a ring outer electrodes creates the radial electric field, which endows the exciton with a dipole moment $\mathbf{d}$ in a direction perpendicular to its circular motion. The external magnetic field is oriented along the growth axis. This configuration guarantees a mutual orthogonality of the induced electric dipole $\mathbf{d}$, the magnetic field $\mathbf{B}$ and the azimuthal component of the condensate wave vector.

In this section we consider the regime of moderate electric and magnetic fields, when the exciton relative motion wave function is localised mainly in the Coulomb potential well, see  Fig.~\ref{Fig.States}a.

\subsection{Guage field for a single exciton}

The Hamiltonian describing a motion of the interacting electron-hole pair in the crossed electric $\bm{\mathcal{E}}$ and magnetic $\mathbf{B}$ fields in the symmetric gauge reads \cite{Lozovik2002} 
	\begin{eqnarray}\label{Hrel}
		\hat H_{\rm ex} = \frac{{\mathbf{p}}^2}{2M} -\frac{\hbar^2}{2\mu}\Delta_{\mathbf{r}} + i\frac{e\hbar B}{2\nu} \bm{\nabla}_{\mathbf{r}} \times \mathbf{r}+ \frac{e^2 B^2}{8\mu} r^2  \\
		\nonumber + V_c(\mathbf{r}) - e \mathbf{r} \bm{\mathcal{E}}  + e\mathbf{r}\frac{\mathbf{B} \times {\mathbf{p}}}{M} ,
	\end{eqnarray}
where $M=m_e + m_h$ is the exciton mass associated with its center of mass motion, while the relative mass $\mu=m_e m_h\left/{M}\right.$ and the parameter $\nu=m_e m_h\left/\left(m_e - m_h\right)\right.$   are dependent on the asymmetry between the electron $m_e$ and the hole $m_h$ masses; $\mathbf{r}$ is the electron-hole separation vector; $V_c=- \varkappa e^2\left/ \left| \mathbf{r} \right| \right.$ is a Coulomb potential with $\varkappa=(4\pi \varepsilon_0 \varepsilon)^{-1}$. 
The last term in \eqref{Hrel} stands for the coupling between the exciton internal structure and its center of mass momentum \cite{Gorkov1968}. 
The last two terms can be cast as $-e\mathbf{r}\left(\bm{\mathcal{E}}+\bm{\mathcal{E}}_b\right)$, where $\bm{\mathcal{E}}_b= {\mathbf{p}} \times \mathbf{B} / M$ is the effective field.

We note that strong magnetic fields are also expected to induce significant Zeeman splittings. In this work we limit our consideration to single-component polaritons that are devoid of the polarization degree of freedom. In the presence of a strong magnetic field it would correspond to a 100\% circularly polarized polariton flow.

The problem of a ground state of the Hamiltonian \eqref{Hrel} has the rigorous solutions in the case of zero \cite{Pederson2007} or strong magnetic fields \cite{Lerner1980,Lozovik2002}. However in the case where both $\bm{\mathcal{E}}$ and $\mathbf{{B}}$ are non-vanishing the excitonic problem cannot be resolved  analytically. For this reason we approach this problem variationally. 

The trial wave function of the relative motion of the bound electron-hole pair is taken in the form
\begin{equation}\label{ansatz}
	\chi(\mathbf{r})=\sqrt{\frac{2}{\pi}} \frac{(1-\lambda^2)^{3/4}}{a} \exp\left(-\frac{|\mathbf{r}| - \lambda x }{a}\right),
\end{equation}
where the Cartesian $x$-axis is parallel to the electric field direction, $a$ and $\lambda$ are variational parameters. The prefactor of the exponent satisfies the normalization condition $\int |\chi(\mathbf{r})|^2 d\mathbf{r}$=1. The value of $a$ corresponds to the effective in-plane exciton Bohr radius, while $\lambda$ accounts for the exciton polarization due to the electric field.

The minimization procedure for the energy $E_{\rm ex}(\mathbf{k})=\int \hat{H}_{\rm ex} |\chi(\mathbf{r})|^2 d\mathbf{r}$ of the exciton moving with a momentum $\mathbf{p}=\hbar \mathbf{k}$ yields:
\begin{equation}\label{Eex}
	E_{\rm ex}(\mathbf{k}) = \frac{\hbar^2\mathbf{k}^2}{2M} + \frac{\hbar}{2\mu a^2} -\frac{ 2\varkappa e^2}{a} 	+\frac{3e^2B^2a^2}{16\mu}  - \frac{\alpha\left({\bm{\mathcal{E}}}+{\bm{\mathcal{E}}}_b\right)^2}{2},
\end{equation}
where we have assumed that the exciton polarization   is weak, $\lambda^2 \ll 1$. Here the fourth term accounts for the Langevin diamagnetic shift while the last one stands for the DC-Stark shift from the combined action of the real $\bm{\mathcal{E}}$ and effective $\bm{\mathcal{E}}_b$ electric fields. The DC Stark effect if governed by the exciton polarizability which reads:
\begin{equation}\label{polarizability}
	\alpha=\frac{\alpha_0 \tau^3}{1+\alpha_0 \frac{5\tau^3}{12 \mu} B^2},
\end{equation}
where $\alpha_0 = 9/8\varkappa^{-1} a_{\rm B}^3$ defines the 2D exciton polarizability at zero magnetic field  with $a_{\rm B} = {\hbar^2}\left/\left({2\varkappa e^2 \mu}\right)\right.$ being the 2D exciton Bohr radius. 
The dimensionless parameter $\tau(B)=a(B)/a_{\rm B}$ quantifies the reduction of the 2D exciton size in the presence of magnetic field \cite{Kavokin1993}. 

Taking into account that the effective field  $\bm{\mathcal{E}}_b={\hbar} \mathbf{k}\times\mathbf{B}/M $ is linear in the exciton momentum, one can recast equation \eqref{Eex} in the form
\begin{equation}\label{EexGF}
	E_{\rm ex}= E_{\rm hl} + \frac{\left(\hbar\mathbf{k} - \mathbf{A}_{\rm HMW}\right)^2}{2m_{\rm ex}}  - \frac{{A}_{\rm HMW}^2}{m_{\rm ex}},
\end{equation}
where the effective gauge potential 
\begin{equation}\label{Aex}
	\mathbf{A}_{\rm HMW} = e\mathbf{d} \times \mathbf{B}  \frac{m_{\rm ex}}{M}
\end{equation}
is proportional to the dipole moment $-e\mathbf{d}=\alpha \bm{\mathcal{E}}$ induced by the external electric field, {$m_{\rm ex}^{-1} =M^{-1}\left(1- \alpha B^2/M\right)$} is the exciton effective mass renormalized due to magnetoelectric Stark effect. The zero-momentum energy of the hydrogen-like exciton is given by $E_{\rm hl}   =\frac{\hbar}{2\mu a^2} -\frac{ 2\varkappa e^2}{a}+\frac{3e^2B^2a^2}{16\mu}- \frac{\alpha\mathcal{E}}{2}$.

The expression \eqref{Aex} for the effective gauge field for dipole-polarized exciton provides an instructive  insight  into the origin of this phenomenon and its connection with the HMW effect. According to the original proposal of He, McKellar and Wilkens \cite{He1993,Wilkens1994} the phase accumulated by the electric dipole $q \mathbf{d}$ moving on a path $\mathcal{D}$ is $\phi_{\rm HMW} = q\hbar^{-1}\int_\mathcal{D} \mathbf{d} \times \mathbf{B} \cdot d{\bm{l}}$, where $\bm{l}$ is the coordinate along the pass. Thus, the value $\mathbf{A}_{\rm HMW} = q \mathbf{d}\times\mathbf{B}$ can be associated with the effective gauge potential, $\phi_{\rm HMW}=\hbar^{-1}\int_{\mathcal{D}} \mathbf{A}_{\rm HMW} \cdot d\bm{l}$, which obeys the standard pulse rescaling  rule $\mathbf{p} \rightarrow \mathbf{p} - \mathbf{A}_{\rm HMW}$. This expression for $\mathbf{A}_{\rm HMW}$ is identical to Eq.~\eqref{Aex} up to the mass renormalization factor.  Note that the presence of the gauge field for dipole-polarized excitons governed by the Hamiltonian \eqref{Hrel} was pointed out by Sonin in \cite{Sonin2009,Sonin2010}.

According to \eqref{Aex}, the gauge field strength $A_{\rm HMW}$ scales linearly with the polarizing electric field  while its magnetic field dependence is primarily governed by the exciton polarizability~\eqref{polarizability}. Note that at $B=0$ our variational solution only slightly underestimates the 2D exciton polarizability whose exact value is $\alpha_0=21/16\varkappa^{-1} a_{\rm B}^3$ \cite{Pederson2007}.

\begin{figure}
	\includegraphics[width=\linewidth]{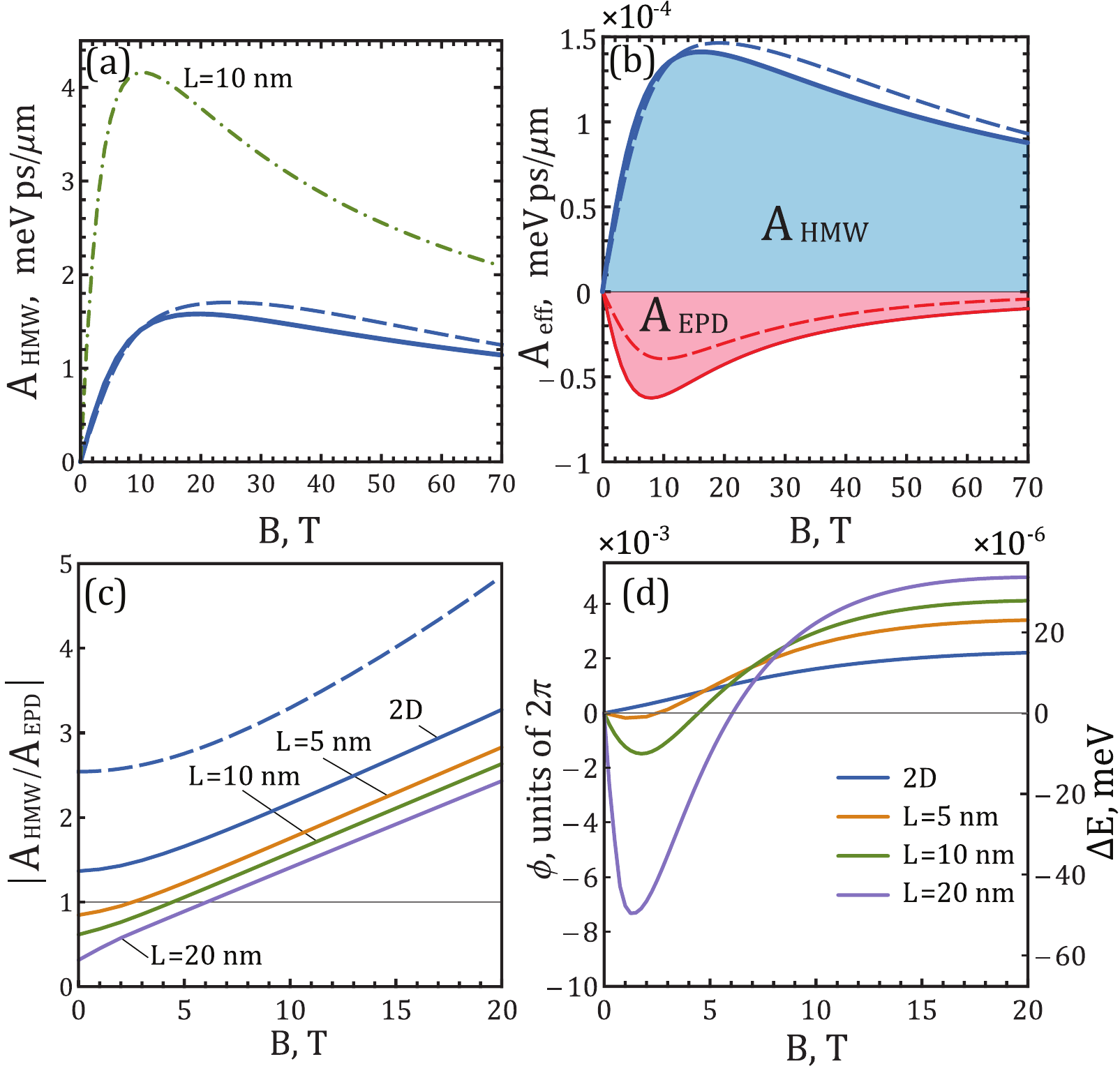} 
	\caption{(a) The dependence of the exciton gauge field strength  $A_{\rm HMW}$  on the magnetic field for the 2D GaAs exciton  (blue lines) at the external electric field $\mathcal{E}=1$~$\rm kV/cm$. The analytical solution \eqref{Aex} is shown with the dashed line while the results extracted from numerical simulations correspond to the solid line. {The dash-dotted line corresponds to the case of the QW well of the thickness $L=10$~nm.} (b) The magnitudes of the polariton gauge potentials. (c) $A_{\rm HMW}/A_{\rm EPD}$ relation for experimentally feasible range of $B$. The dashed line corresponds to Eqs.~\eqref{ApolHMW} and \eqref{ApolEPD} while the solid lines show the data obtained numerically. (d) The geometric phase difference $\phi$ between polaritons travelling in opposite arms of a ring-shaped Berry-phase interferometer. The right axis shows  the energy splitting between the clockwise and the anti-clockwise currents of polariton superfluid excited on the thin ring, see Fig.~\ref{fig1}. The ring radius is $R_0=15$~$\mu$m. The material parameters under consideration are typical for the GaAs medium: $m_e=0.067m_{\rm fe}$, $m_h=0.45m_{\rm fe}$,  $\varepsilon = 12.9$ where $m_{\rm fe}$ is the free-electron mass.} \label{fig.analytics}
\end{figure}

At weak magnetic fields $B<\sqrt{12\mu \alpha_0/5}$ and $\tau\approx 1$, the exciton polarizability is approximately constant and the gauge field grows linearly with the magnetic field \cite{Kozin2020}, see Fig.~\ref{fig.analytics}a. At the same time in the limit of strong magnetic field  polarizability decreases as $\alpha\propto B^{-2}$ and then $A_{\rm HMW} \propto B^{-1}$. This behaviour can be connected with the exciton size reduction induced by the magnetic field \cite{Kavokin1993} {and the gradual transition to the magnetoexciton regime \cite{Lerner1980}}. The maximum value of the gauge field strength is achieved in the intermediate regime. {For the considered parameters which are typical for GaAs structures, the magnetic field that maximizes the gauge field would be about 20 T for the case of an idealized 2D exciton. In realistic quantum wells (QWs) the exciton radius is larger than one of a 2D exciton and thus the polarizability \eqref{polarizability} decreases faster as the magnetic field grows. In particular, for GaAs QW of $L=10$~nm thickness the gauge field maximum corresponds to $B\approx 10$~T, see Fig.~\ref{fig.analytics}a.} 

To check the validity of our variational approach \eqref{ansatz}, we calculate the ground state of the Hamiltonian \eqref{Hrel} numerically using the finite element method (FEM). The corresponding gauge potential is shown in Fig.~\ref{fig.analytics}a with the solid line. The results of simulations demonstrate a high degree of reliability of our variational approach.

\subsection{Gauge field for exciton polaritons}

In the regime of strong light-matter coupling, the excitons impart their sensitivity to the effective gauge field to polaritons. The interaction of the exciton with the microcavity mode can be described by the model of two coupled oscillators. It yields that the energy of the lower branch polariton reads
\begin{equation}\label{ELP}
	E_{\rm LP} = \frac{1}{2} \left( E_{\rm ex}+ E_{\rm ph} - \sqrt{\left[ E_{\rm ex} - E_{\rm ph} \right]^2+ \Omega^2} \right),
\end{equation}
where the microcavity mode dispersion is $E_{\rm ph} = E_{\rm{c}} + \hbar^2 k^2 \left/2m_{\rm ph} \right.$, $\Omega$ is the exciton-photon coupling strength (resonant Rabi splitting) and $m_{\rm ph}$ is the microcavity photon mass.  For the sake of self-consistency, we  add the semiconductor band gap energy $E_g$ to the exciton energy \eqref{EexGF}.

According to its definition, the Rabi splitting $\Omega$ is dependent on the electron-hole relative motion wave function taken at the coincidence of the electron and hole positions, $\Omega \propto \chi(0)$. With the trial function \eqref{ansatz} we have $\chi(0)=\sqrt{2/\pi a^2} \left(1-\lambda^2\right)^{3/4}$. Then assuming $\lambda^2 \ll 1$, we expand the coupling strength to the first order in $\lambda^2$:
\begin{equation}\label{RabiApproximation}
	\Omega \approx (1-3/4\lambda^2)\Omega_0,
\end{equation}
where $\Omega_0 \equiv \Omega_0(B)$ defines the coupling strength for a non-polarized exciton at $\bm{\mathcal{E}}=0$.
Minimization of the exciton energy thus yields  $\lambda=\lambda_0 (\mathcal{E}+\mathcal{E}_b)$, where
\begin{equation}\label{lambda0}
	\lambda_0=-\frac{3a_{\rm B}^2}{4e\varkappa}\frac{\tau^2}{1+\alpha_0 \frac{5\tau^3}{12 \mu} B^2}.
\end{equation}

We assume that all the momentum-dependent terms under the square root in \eqref{ELP} are small compared to the Rabi splitting $\Omega_R=\sqrt{\Delta^2 + \Omega_0^2}$, where $\Delta= E_g+ E_{\rm hl} - E_c$. Then after a simple algebra one obtains:
\begin{equation}
	E_{\rm LP} = E_{\rm LP}^{(0)} + \frac{\left(\hbar \mathbf{k} - \mathbf{A}_{\rm pol}\right)^2}{2m_{\rm pol}} - \frac{A_{\rm pol}^2}{2m_{\rm pol}}.
\end{equation}
Here $E_{\rm LP}^{(0)}=1/2\left(E_g+E_{\rm hl} +E_c - \Omega_R\right)$, the polariton mass is {$m_{\rm pol}^{-1} = C_x^2 m_{\rm ex}^{-1}+C_p^2m_{\rm ph}^{-1} - \frac{9\Omega_0^2}{8\Omega_R}B^2\lambda_0^4\mathcal{E}^2M^{-2}$} and $C_{x,p}=2^{-1/2}\left(1 \mp \Delta\left/\Omega_R\right.\right)^{1/2}$ are the Hopfield coefficients. The effective gauge potential accounts for the contribution from two effects:
\begin{equation}\label{Apol}
	\mathbf{A}_{\rm pol} = \mathbf{A}_{\rm HMW} + \mathbf{A}_{\rm EPD},
\end{equation}
where the first term 
\begin{subequations}\label{ApolDetailed}
	\begin{equation}\label{ApolHMW}
		\mathbf{A}_{\rm HMW} =C_x^2 e\mathbf{d}\times\mathbf{B}\frac{m_{\rm pol}}{M}
	\end{equation}
corresponds to the gauge field associated with the HMW phase. It is crucial that the gauge field strength for polaritons is reduced compared to the excitons \eqref{Aex} by the factor of $m_{\rm pol}/M$ which is typically of the order of $10^{-4}$. This is because the HMW gauge field originates from the motion of excitons.  Polaritons in contrast to excitons are light quasiparticles, which inherit their transport properties mainly from their photonic part. Therefore, the exciton gauge field has a suppressed impact on the  effective dynamics of polaritons, see Figs.~\ref{fig.analytics}a,b. 

The second term in \eqref{Apol} describes the gauge field arising due to the dipole-polarization-induced decoupling of the exciton and the microcavity mode:
\begin{equation}\label{ApolEPD}
	\mathbf{A}_{\rm EPD} \simeq -\frac{3\Omega_0^2}{4\Omega_R}\lambda_0^2  {\mathbf{B}\times\bm{\mathcal{E}}}  \frac{m_{\rm pol}}{M}.
\end{equation}
\end{subequations}
Note that vectors $\mathbf{A}_{\rm HMW}$ and $\mathbf{A}_{\rm EPD}$ are anti-collinear. Indeed, the reduction of a coupling strength increases polariton energy \eqref{ELP}, while the $\mathbf{k}$-dependent DC Stark shift of the exciton level is always negative, Eq.~\eqref{Eex}. So, the strength of the resulting artificial gauge field \eqref{Apol} is determined by the competition of these effects.  At the weak magnetic field, $\tau \approx 1$, and  zero exciton-photon detuning $\Delta=0$, Eqs.~\eqref{ApolHMW} and \eqref{ApolEPD} give the ratio:
\begin{equation}\label{RatioAnal}
\left|\frac{{A}_{\rm HMW}}{{A}_{\rm EPD}}\right| =\frac{4E_{b0}}{3\Omega_0},
\end{equation} 
{where $E_{b0} =2\varkappa^2 e^4\mu \hbar^{-2}$ stands for the 2D exciton binding energy at zero magnetic field. For GaAs-based microcavities considered here, the Rabi splitting can be as high as $10$~meV while the 2D exciton binding energy is $E_{b0} \simeq 19$~meV. In this case the effect associated with the HMW phase dominates over the mechanism of Rabi splitting renormalization, see Fig.~\ref{fig.analytics}b. The ratio $|{A}_{\rm HMW}/{A}_{\rm EPD}|$ in the experimentally feasible range of magnetic field $B$ is shown in Fig.~\ref{fig.analytics}c. The imbalance between the gauge potentials gradually grows as the magnetic field increases.  }

The numerical solution for the 2D exciton ground state (the solid lines in Fig.~\ref{fig.analytics}b) demonstrates that our variational ansatz  underestimates the effect of the exciton  deformation responsible for the Rabi-splitting renormalization. The actual ratio $\left|{{A}_{\rm HMW}}\left/{{A}_{\rm EPD}}\right.\right|$ is close to unity at weak magnetic field  when both mechanisms responsible for the gauge field generation must be taken into account, see Fig.~\ref{fig.analytics}c. 

{In QWs the exciton binding energy is significantly smaller than the two-dimensional exciton binding energy because of the finite widths of the QWs that allows for a significant spread of electron and hole envelope functions in normal to the QW plane direction. The exciton radius increases with the increase of the QW width, which leads to the decrease in  the exciton-photon coupling strength. However, the Rabi-splitting reduction can be compensated by an increase of the number of QWs embedded in the microcavity. As a result, the Rabi splitting in GaAs-based microcavities can be comparable with the exciton binding energy \cite{Askitopoulos2018}. In this case the Rabi-splitting renormalization  dominates at weak magnetic field. It is demonstrated in Fig.~\ref{fig.analytics}c which shows the ratio $\left|{{A}_{\rm HMW}}\left/{{A}_{\rm EPD}}\right.\right|$  for QWs of thicknesses $L=5,\, 10 $ and $20$~nm for which the exciton binding energy is $13$, $10.6$ and $8$ meV,  respectively. Here we use the same value of the  Rabi splitting at zero magnetic and electric fields $\Omega_0(B=0)=10$~meV. At strong magnetic field  the gauge potential coming from the HMW effect dominates in accordance with the predictions of our variational approach.}

The solutions for the QW excitons were found numerically as the eigenstates of the Hamiltonian \eqref{Hrel} with the effective Coulomb potential $V_c(r) = -e^2 \varkappa\iint |v_e(z_e)|^2|v_h(z_h)|^2\left/ \sqrt{(z_e-z_h)^2+r^2}\right. dz_e dz_h$. Here the electron and hole wave functions in the growth direction were taken as $v_{e,h}(z_{e,h})=\sqrt{2/L}\sin\left(z_{e,h}\pi/L\right)$.

Several strategies for detecting the presence of the polariton gauge field in a ring geometry can be applied. In the case of a ring-shaped polariton condensate shown in Fig.~\ref{fig1}, the radial electric field creates the angular gauge potential $\mathbf{A}_{\rm pol}={A}_{\rm pol}\hat{\varphi}$, where $\hat{\varphi}$ is the polar angle. It imposes chirality on the polariton superfluid in a sense of the energetically favourable direction for a circular polariton flow. Therefore, in the case of polariton condensate localised on a thin ring with the radius $R_0$, the energies of the counter-propagating circular currents with unit vorticities are split by
\begin{equation}
	\Delta E=\frac{2\hbar A_{\rm pol}}{m_{\rm pol}R_0}.
\end{equation}
Due to the strong imbalance between the exciton and polariton masses discussed above, the gauge field is weak. So, the energy splitting $\Delta E$ turns out to be very low for the feasible experimental parameters, see Fig.~\ref{fig.analytics}d.
In particular, even if the electric field is as high as several {kV$/$cm}, the splitting remains below $10^{-4}$~meV. It is expected to be hidden behind the spectral broadening of polariton states which is about tens of $\mu$eV \cite{Askitopoulos2018}. Increasing the splitting above this value requires the electric fields about $10^3$~kV$/$cm at which the exciton dissociates.

{Note that due to the interplay between the HMW effect and the Rabi-splitting renormalization QWs, the effective gauge field changes its sign as the magnetic field increases, see Fig.~\ref{fig.analytics}d. This effect is pronounced in the wide QWs for which  the Rabi splitting exceeds the binding energy.}

The alternative approach consists in using of a ring Berry-phase interferometer  \cite{Shelykh2009}  designed in such a way that interferometer arms encircle the radial electric field divergency point. In this case the gauge potential is responsible for the  difference of geometrical phases gained by polaritons travelling in different arms. % along the circle of a total length $\oint_\mathcal{C}d\bm{l}=2\pi R_0$
The resulted phase difference is 
\begin{equation}\label{BerryPhase}
	\phi=\hbar^{-1}\oint \mathbf{A_{\rm pol}}\cdot d\bm{l} = 2\pi R_0\hbar^{-1}A_{\rm pol}.
\end{equation}
While the interference effects  are pronounced at $\phi \sim \pi$, the expecting geometric phase accumulated in the polariton ring interferometer is smaller by two orders of magnitude even for the path circumference as high as few hundred micrometers, see Fig.~\ref{fig.analytics}d.

\section{Crossover between the hydrogenic and the dipole-polarized states}\label{Sec.III}

\subsection{ Synthetic gauge field for polaritons formed by a non-hydrogen-like exciton}\label{Sec.IIIA}
Since the $\mathbf{A}_{\rm pol}(B)$-dependence has a maximum corresponding to the optimal magnetic field, the only remaining strategy for increasing of the gauge field strength is to amplify the polarizing electric field $\bm{\mathcal{E}}$. However the strong electric field substantially modifies the exciton internal structure, so the  variational approach based on the hydrogen-like exciton wave function \eqref{ansatz} fails. In this regime, the effect of polarizing field can not be treated perturbatively and the expressions \eqref{Eex}, \eqref{EexGF} and \eqref{Aex} do not describe  the linear-in-$\mathbf{k}$ response of the exciton energy. 

For this reason, in the non-hydrogen exciton regime, the required linear-in-$\mathbf{k}$ dependencies of the exciton energy and the Rabi splitting can be extracted from the expansion of $E_{\rm ex}(\mathcal{E})$- and  $\Omega(\mathcal{E})$-dependencies around the external electric field $\mathcal{E}$:
\begin{subequations}
	\begin{eqnarray}\label{expansion}
		E_{\rm ex}(\mathcal{E}+\mathcal{E}_b) &\simeq& E_{\rm ex}(\mathcal{E}) + \mathcal{E}_b \frac{\partial E_{\rm ex}}{\partial \mathcal{E}}  + \mathcal{O}\left[\mathcal{E}_b^2\right],\label{expansion1}
		\\
		\xi(\mathcal{E}+\mathcal{E}_b) &\simeq& \xi(\mathcal{E}) + \mathcal{E}_b \frac{\partial \xi}{\partial \mathcal{E}}  + \mathcal{O}\left[\mathcal{E}_b^2\right], \label{expansion2}
	\end{eqnarray} 
\end{subequations}
where $\xi=\Omega/\Omega_0$ quantifies the coupling strength reduction. This approach takes advantage from the notion that the artificial gauge field stems from the modification of the exciton's energy and internal structure by the  effective electric field $\bm{\mathcal{E}}_b$ associated with its motion. For the exciton states with small momenta situated within the light cone and at feasible magnetic field, the effective field is typically weak  and does not exceed $0.1$~$\rm kV/cm$. Therefore, one can consider ${\mathcal{E}}_b$ as a small parameter $\mathcal{E}_b \ll \mathcal{E}_0$ of the expansions \eqref{expansion1} and \eqref{expansion2}. 

The factors at the terms linear in effective field in \eqref{expansion1} and \eqref{expansion2} can be associated with the gauge fields magnitudes. Following the procedure from the previous section, one obtains:
\begin{subequations}\label{NHLE_GaugeField}
	\begin{eqnarray}
{A}_{\rm HMW}&=& - C_x^2 \frac{\partial E_{\rm ex}}{\partial{\mathcal{E}}}B \frac{m_{\rm pol}}{M}, \label{NHLE_HMW}
\\
{A}_{\rm EPD}&=&  \frac{\Omega_0^2}{4\Omega_R}\frac{\partial \xi^2}{\partial{\mathcal{E}}}B \frac{m_{\rm pol}}{M} \label{NHLE_EPD},
\end{eqnarray}
\end{subequations}
where the polariton effective mass is $m_{\rm pol}^{-1} = C_x^2 m_{\rm ex}^{-1}+C_p^2m_{\rm ph}^{-1} - \frac{\Omega_0^2}{2\Omega_R}\left(\partial \xi/\partial \mathcal{E}\right)^2 B^2/M^{2}$.
Note that these expressions are consistent with Eqs.~\eqref{ApolHMW} and \eqref{ApolEPD} once $E_{\rm ex}$ is given by  \eqref{EexGF} and $\xi$ is defined by Eqs.~\eqref{RabiApproximation} and \eqref{lambda0}. Therefore, the polariton gauge field magnitude can be connected with the sensitivity of the exciton energy and the Rabi splitting to the electric field strength variation. In the next section, we demonstrate that this sensitivity can be sufficiently increased when the exciton state experiences a crossover to the strongly decentred solution.

\subsection{The crossover regime}

The conditions for the crossover existence can be determined using the effective potential representation. The terms accounting for the diamagnetic shift, attractive electron-hole interaction and the  electric field in the Hamiltonian \eqref{Hrel}  can be regrouped as $W(\mathbf{r}) = V_d(\mathbf{r})+V_c(\mathbf{r})$, where
\begin{equation}\label{DoubleWell}
	V_d(\mathbf{r}) = \frac{\mu\omega_c^2}{8} (\mathbf{r} + \bm{\rho}_0)^2 - \frac{  e \bm{\rho}_0 \bm{\mathcal{E}}}{2}  
\end{equation}
stands for the parabolic diamagnetic potential shifted  along the electric field direction by $\bm{\rho}_0= 4\mu \bm{\mathcal{E}}/(eB^2)$. Here $\omega_c = eB/\mu$ is the electron cyclotron frequency. The negative energy offset $-e \bm{\rho}_0 \bm{\mathcal{E}}/{2}$ is caused by the presence of electric field.

In the absence of any external field the electron-hole relative motion is bounded in the Coulomb potential well $V_c(\mathbf{r})$. This is the case of the HLE whose energy $E_{\rm hl}$ and structure are described by Eqs.~\eqref{EexGF} and \eqref{ansatz}, respectively. When the shift $\rho_0$ of the diamagnetic potential position is small, i.e. at weak electric and strong magnetic fields, the Coulomb $V_c$ and the diamagnetic $V_d$ potentials coalesce, see Fig.~\ref{Fig.States}a. In this case magnetic field is responsible for the squeezing of the exciton wave function \cite{Kavokin1993}, while the electric field has a merely perturbative effect causing  formation of the dipole-polarized exciton.

At strong electric and weak magnetic field so that the shift  $\rho_0$ of the diamagnetic well position is large, the total potential $W(\mathbf{r})$ has a   double-well shape, Fig.~\ref{Fig.States}c. In this case each potential well can be characterized by the independent set of energy levels. For the states localized close to the diamagnetic well minimum, the presence of the Coulomb interaction can be accounted perturbatively. These states thus resemble the states of a free particle in the magnetic field  \cite{Lozovik2002} with the characteristic localization scale equal to the magnetic length $l_c=\sqrt{\hbar/(eB)}$ and with the energy $E_d \approx \hbar \omega_c /2 - e{\rho}_0 {\mathcal{E}}/2 - \varkappa e^2/\left|{\rho_0}\right|$. Note that the negative energy offset in this case accounts for the interaction of the dipole $e \rho_0$ with the electric field. For large $\rho_0$ it typically means that the lowest level of the diamagnetic potential  lies below the hydrogen-like exciton and corresponds to the ground state. However, because of the strongly decentred relative electron-hole position, this state does not couple to the photon field. The polaritons are formed by the hydrogenic excitons. Due to the vanishing overlap between the states located in the potential minima at $\rho_0 \gg \{a, l_c\}$, the HLE state remains unaffected by the presence of the second potential well.

\begin{figure}
	\includegraphics[width=\linewidth]{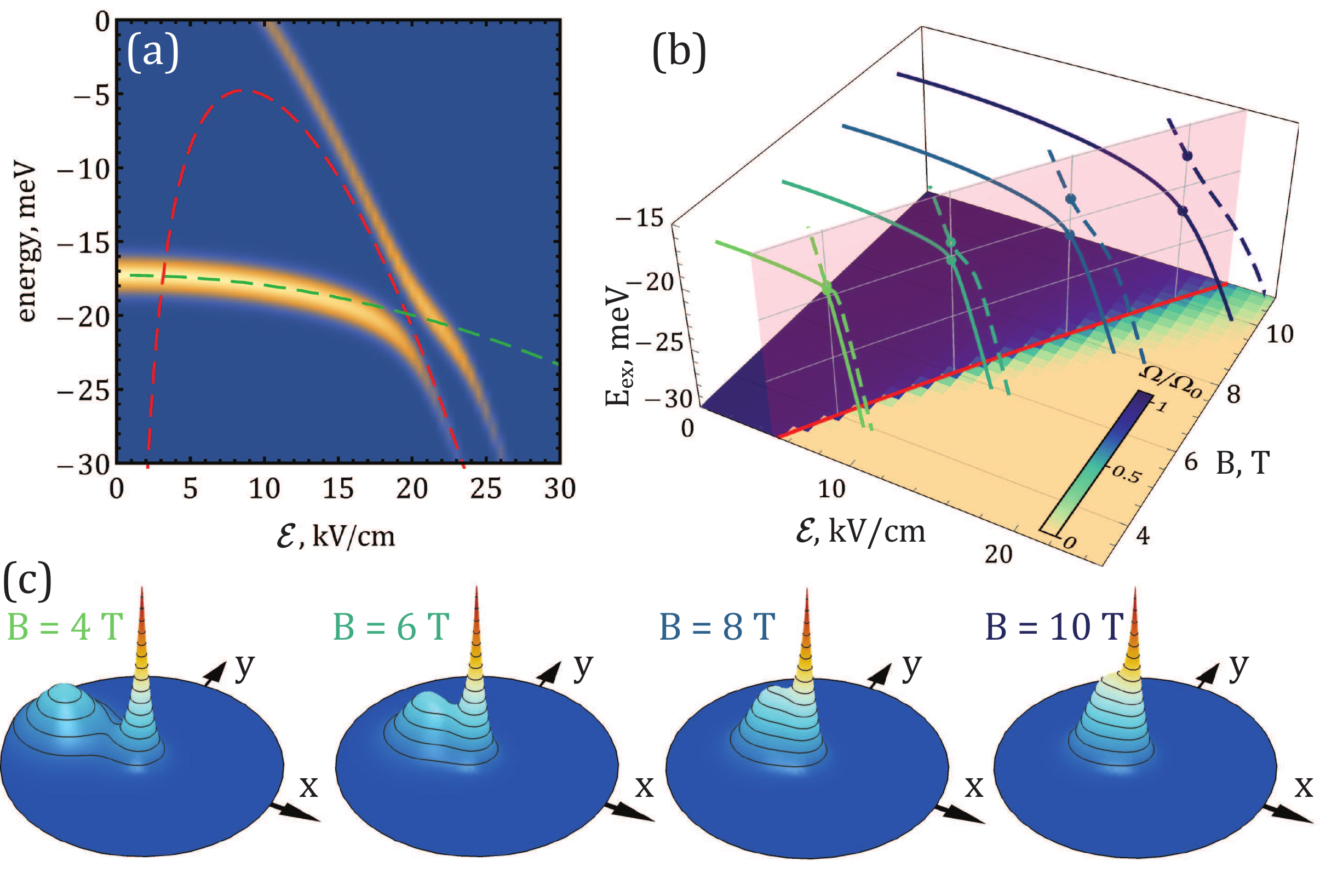}
	\caption{
(a) Anticrossing of two least energy exciton states at $B=10$~T calculated numerically. The false color demonstrates the effect of the Rabi-splitting renormalization. The line brightness encodes the exciton oscillator strength variation quantified by the parameter $\xi=\Omega/\Omega_0$. The green dashed line shows the energy of the hydrogen-like exciton \eqref{EexGF} at zero momentum, while the red dashed curve corresponds to the least energy state of the diamagnetic potential well, $E_d$.
(b) The anticrossing of the  energy levels  for different values of the magnetic field. At the bottom, the effect of exciton oscillator strength reduction at the crossover is illustrated. The red facet indicates the position of the crossover estimated by equating $E_d$ and $E_{\rm hl}$. (c) The relative motion wave function $|\chi(\mathbf{r})|^2$ of the lowest energy state in the crossover regime at four different values of the magnetic field shown in panel (b).
}\label{Fig.Crossover}
\end{figure}

For the intermediated values of the shift parameter $\rho_0$, the minima of the effective potential $W(\mathbf{r})$ get closer to each other so the tunnelling through the barrier between the potential wells becomes possible, see Fig.~\ref{Fig.States}b. This leads to the mixing of the HLE and the strongly polarized states. The crossover manifests itself in the anticrossing in the exciton energy spectra. In particular, Fig.~\ref{Fig.Crossover}a demonstrates the numerically calculated energies of the two least energy states as a function of electric field at $B=10$~T. The splitting of the energy levels occurs at the point at which the energy of the decentred state $E_d$ (red dashed curve) is equal to the HLE energy $E_{\rm hl}$ (green dashed line). This condition provides a simple estimate of the crossover position. 

The energy of the splitting is governed by the overlap of the relative wave functions. As the magnetic field decreases, the inter-well separation $\rho_0$  grows gradually and the splitting at the crossover position weakens as it is shown in Fig.~\ref{Fig.Crossover}b. As a result, at weak magnetic field (below $B \approx 3$~T for the considered material parameters), the anticrossing disappears since the potential wells are separated too far for providing an efficient coupling between them. The effect of separation of the effective potential wells is illustrated in  Fig.~\ref{Fig.Crossover}c where the structure of the lowest energy states at the crossover position is shown at four values of the magnetic field. 

The crossover from the hydrogenic to the dipole-polarized state is accompanied by a significant modification of the exciton oscillator strength which governs the value of the polariton Rabi splitting, see Fig.~\ref{Fig.Crossover}a. This is because the hydrogenic wave function is pushed out form the Coulomb well as the electric field increases beyond the region of anticrossing. Eventually it is superposed by the decentred state which is almost decoupled from the photon field. This transition is illustrated in Fig.~\ref{Fig.Crossover}b where the variation of ground state oscillator strength is shown in the parameter plane $(\mathcal{E},B)$. The regions of the weak and strong oscillator strength of the lowest energy exciton state are separated by the red line indicating the crossover position. This boundary is obtained by equating the energies of the  dipole-polarized  state $E_d$  and the zero-momentum hydrogen-like exciton $E_{\rm hl}$. Note that the oscillator strength of the  upper state gradually grows due to the mixing with the hydrogenic exciton states as the electric field approaches the anticrossing region. This important feature of the crossover regime affects the properties of the polariton states as it will be considered in the next section.

\section{The giant artificial gauge field for polaritons in the crossover regime}\label{Sec.IV}

According to Eqs.~\eqref{NHLE_HMW} and  \eqref{NHLE_EPD}, the magnitude of the polariton artificial gauge field is determined by the slope of  the $E(\mathcal{E})$- and $\Omega(\mathcal{E})$-dependencies. For both the slope increases in the vicinity of the crossover to the strongly polarized exciton state. Thus one can expect a significant enhancement of the gauge field strength in this parameter domain. However, in the vicinity of the   anticrossing  the optical response of the medium is determined by the presence of several exciton resonances. In this case, the polariton energies can be  found by the solution of the eigenvalue problem for the following three-mode Hamiltonian:
\begin{equation}\label{H3x3}
	\hat{H}=\left(
	\begin{array}{ccc}
		E_{\rm ex}^{(1)}(\mathbf{k}) & 0 &\Omega_1/2 \\
		0 & E_{\rm ex}^{(2)}(\mathbf{k}) & \Omega_2/2  \\
	    \Omega_1/2 & \Omega_2/2 &  E_{\rm ph}(\mathbf{k})
	\end{array}
	\right),
\end{equation}
where $E_{\rm ex}^{(1,2)}$ and $\Omega_{1,2}$ stand for the energies and the exciton-photon coupling strengths for the two lowest exciton states. Here we neglected by the presence of  the upper exciton states since their oscillator strengths are typically weak near the crossover. The diagonalization of the Hamiltonian \eqref{H3x3} yields three polariton branches. We focus on the lowest polariton state whose energy is denoted as $E_{\rm pol}$.

The polariton artificial gauge field can be determined by analogy with the approach used in  Sec~.\ref{Sec.IIIA}: 
\begin{equation}\label{Apol3modes}
	\mathbf{A}_{\rm pol} = - \frac{\partial E_{\rm pol}}{\partial{\mathcal{E}}}B \frac{m_{\rm pol}}{M},
\end{equation}
Note that Eq.~\eqref{Apol3modes} accounts simultaneously for the gauge field arising from the Rabi-splitting renormalization and the one associated with the HMW phase. In the two-mode approximation, $\Omega_2\approx 0$ and $|E_{\rm ex}^{(2)}-E_{\rm ph}| \gg \Omega_1$, \eqref{Apol3modes} reduces to the definitions \eqref{NHLE_HMW} and  \eqref{NHLE_EPD}. 

\begin{figure}
	\includegraphics[width= 0.6 \linewidth]{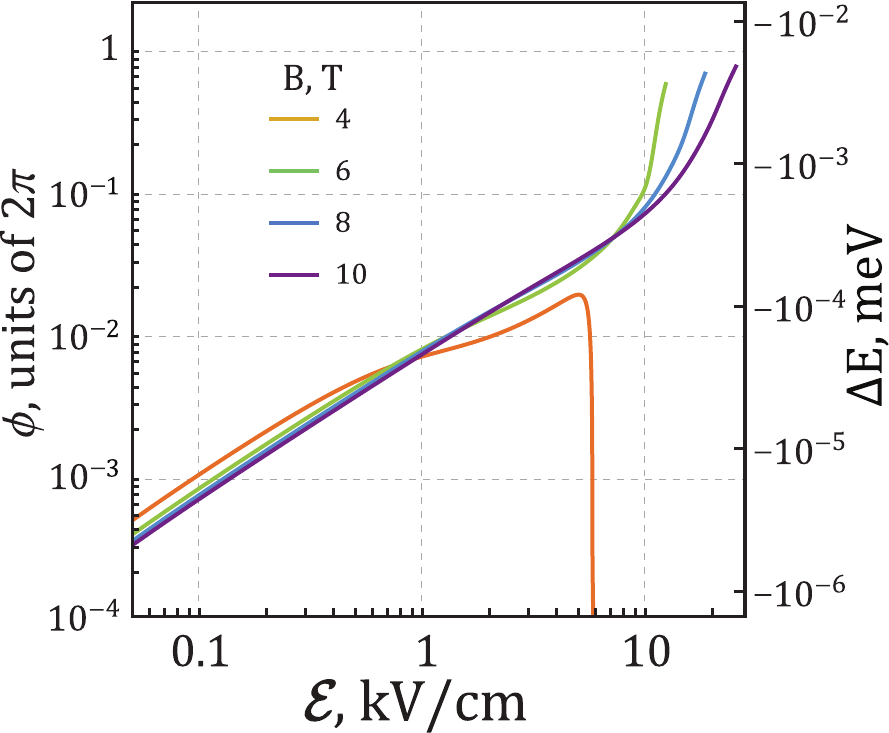}
	\caption{The Berry phase difference gained by polaritons travelling in opposite directions in the ring-shaped  interferometer with the radius $R_0=15$~$\mu$m. The exciton is formed in the 10 nm-thick QW. The polariton Rabi coupling strength for a non-polarized exciton at zero magnetic field is $\Omega_0(B=0)=5$~meV for both panels. {The bare cavity frequency $E_c$ is resonant with the lower exciton state energy.} The energy splitting between the clockwise and the anti-clockwise currents of a polariton condensate localised on the ring is shown in the right axis. {Here for simplicity we neglected the weak dependence of the lower-branch polariton mass  $m_{\rm pol}$ on the electric field.}}\label{Fig.5}
\end{figure}

The enhancement of the polariton gauge field strength in the crossover region is demonstrated in  Fig.~\ref{Fig.5}, which shows the geometric phase $\phi$ gained by polaritons in the ring  Berry-phase interferometer, see \eqref{BerryPhase}, as a function of the polarising electric field. In the HLE regime (below few kV$/$cm for the considered case) the gauge field grows linearly with the increase of the electric field strength as it is described in  \ref{Sec.HLEGaugeField}. In this regime the Berry-phase difference remains below hundredth of $2\pi$. Typically, it does not not allow for the observation of the effects connected with the geometric phase. However, the value of $\phi$ increases sharply  as the polarizing electric field strength approaches the crossover.  

For {the ring GaAs-based interferometer of $R_0=15$~$\mu$m radius}, the Berry phase approaches $2\pi$ before the lowest polariton state becomes strongly exciton-like and decouples from the external light. {In this case the artificial gauge field  can generate the effective magnetic field whose flux $\Phi=e^{-1}\oint \mathbf{A}_{\rm pol}\cdot d\bm{l}$ is close to the flux quantum $\Phi_0=h/e$. This is the value required for the observation of the quantum interference effects  \cite{Wood2016} in the   polariton condensate localised on a ring of the respective radius.}

Note that when the Rabi splitting is small in comparison to the exciton binding energy (see Eq.~\eqref{RatioAnal}), the gauge field associated with the HMW phase dominates over the field arising from the exciton-photon coupling strength renormalization. This is the case shown in Fig.~\ref{Fig.5}. The mixing of the exciton states in the crossover region endows the optically active exciton with the large dipole moment typical for the decentred state. Therefore the HMW phase \eqref{ApolHMW} which polaritons accumulate via their excitonic component  increases drastically. The concurrent decrease of the oscillator strength causes an opposite effect whose strength scales linearly in the Rabi splitting $\Omega_0$ according to \eqref{NHLE_HMW} and  \eqref{NHLE_EPD}. So, for small $\Omega_0$ and a smooth crossover, this effect is suppressed. 

When the overlap between the exciton states localized in the different potential minima is small, the effect of mixing is weak as well as the impact of the HMW effect. In this case the crossover occurs in the narrow region of $\mathcal{E}$ as it shown in Fig.~\ref{Fig.Crossover}b,c for the case of $B=4$~T. It means that the exciton oscillator strength is strongly sensitive to the value of the effective electric field associated with the exciton motion. Therefore, the impact of the Rabi-coupling renormalization effect is strong. In this case the magnitude of the resulting gauge field decreases, see the orange curve in Fig.~\ref{Fig.5} corresponding to $B=4$~T.  

{Our numerical simulations demonstrate that for the Rabi splittings which are comparable or larger than the exciton binding energy, the effect of exciton-photon decoupling can dominate over the HMW effect. However for the GaAs based structures in which the condition $\Omega_0 \lesssim E_{b0}$ is typically valid, the strongest value of the resulted gauge field magnitude can be obtained at the moderate and  weak Rabi splittings, where the gauge field is mainly governed by the HMW effect. The results shown in Fig.~\ref{Fig.5} correspond to $\Omega=5$~meV while for larger Rabi splittings the competition between the two sources of the gauge field  leads to the reductions of the resulted gauge potential.}

\section{Conclusion}

The synthesis of the artificial gauge potential for polaritons in crossed electric and magnetic fields is studied. {We demonstrate that for the hydrogen-like exciton the gauge field associated with the HMW phase typically dominates over the contribution from the exciton-photon decoupling in the structures for which the exciton  binding energy exceeds the Rabi coupling.} However, in the system where the value of the polariton Rabi splitting becomes comparable to the binding energy of the unperturbed exciton, the impact of the Rabi splitting renormalization is significant. In this case the reduction of the exciton oscillator strength leads to the screening of the HMW gauge field potential. Accounting for both mechanisms is thus crucial for the efficient generation of the gauge field for exciton polaritons.

{Besides, we propose the approach which substantially enhances the efficiency of the gauge field synthesis. It implies the use of the partially decentred excitons which combine the large electric dipole with the strong coupling to the photon field. In this regime, the gauge field can generate the flux approaching the flux quantum for the ring-shaped polariton condensate having a radius of several tens of micrometers \cite{Mukherjee2019}. This is the necessary prerequisite for observation of the quantum interference effects such as the gauge-field induced formation of the persistent current \cite{Wood2016}.}

Such a strong synthetic gauge field can be used for the fast coherent manipulation of the polariton condensate excited in the ring geometry. This particular problem is important for realization of the logic operation with macroscopic polariton currents \cite{Xue2019}. The artificial gauge field causes the energy splitting $\Delta E$ between the contra-rotating polariton flows. It approaches tens of $\mu$eV in the crossover regime, see Fig.~\ref{Fig.5}. This is of the same order with the values of the splitting between the level of spatial quantization in the annular optical traps \cite{Askitopoulos2018}. The predicted enhancement of the gauge field synthesis paves the way to the realization and investigation of the new topological phases of matter with exciton polaritons.

\begin{acknowledgments}
We thank Project No. 041020100118 supported by Westlake University and Program 2018R01002 supported by Leading Innovative and Entrepreneur Team Introduction Program of Zhejiang. I.C. also acknowledges funding from NSFC (Grant No. 12050410250). The support from RFBR grants 21-52-10005, 20-02-00515, from the Grant of the President of the Russian Federation for state support of young Russian scientists No. MK-5318.2021.1.2 and from the state task for VlSU in the scientific activity project 0635-2020-0013 are acknowledged.
\end{acknowledgments}

\appendix
\section*{References}

%\begin{thebibliography}{99}

%\end{thebibliography}
%\bibliographystyle{achemso}
%\bibliographystyle{iopart-num}
\bibliography{Synthetic_gauge_field}

\end{document}